\documentclass[twocolumn,amsmath,amssymb,floatfix]{revtex4}
\usepackage{graphicx}% Include figure files
\usepackage{bm}% bold math
\usepackage[usenames]{color}

\newcommand{\beq}{\begin{eqnarray}}
\newcommand{\eeq}{\end{eqnarray}}

\def\der#1#2{\frac{\partial #1}{\partial #2}}

\def \ket#1{|#1\rangle}

\def \be{\begin{equation}}
\def \ee{\end{equation}}
\def \ba{\begin{array}}
\def \ea{\end{array}}
\def \bea{\begin{eqnarray}}
\def \eea{\end{eqnarray}}

\def \l{\left}
\def \r{\right}
\def \rr{\right}
\def \half{{1\over 2}}

\def \H{{\cal{H}}}

\def \W{{\Omega}}

\def \a{{\alpha}}

\def \b{{\beta}}

\def \w{{\omega}}
\def \s{s}
\def \p{p}
\def \q{q}
\def \f{{\varphi}}
\def \x{{\chi}}

\def \G{{\Gamma}}
\def \z{{\zeta}}

\def \on{\overline{n}}
\def \intt{\int\limits}

\def\O {\Omega}
\def \summ{\sum\limits}
\def\intt{\int\limits}
\def\nbar{\overline{n}}

\begin{document}

\title{Superfluid-insulator transition of disordered bosons in one-dimension.}

\author{Ehud Altman$^1$, Yariv Kafri$^2$, Anatoli Polkovnikov$^3$,
Gil Refael$^4$\\
{$^1$\small \em Department of Condensed Matter Physics, The Weizmann
Institute of Science Rehovot, 76100, Israel}\\
{$^2$\small \em Department of Physics, Technion, Haifa 32000,
Israel}\\
{$^3$\small \em Department of Physics, Boston University, Boston,
MA 02215}\\
{$^4$\small \em Dept. of Physics, California Institute of
Technology, MC 114-36, Pasadena, CA 91125}}

\begin{abstract}

We study the superfluid-insulator transition in a one dimensional system
of interacting bosons, modeled as a disordered Josephson array, using
a strong randomness real space renormalization group technique.
Unlike perturbative methods, this
approach does not suffer from run-away flows and allows us to study
the complete phase diagram. We
show that the superfluid insulator transition is always Kosterlitz-
Thouless like in the way that length and time scales diverge at the critical point.
Interestingly however, we find that the transition at strong disorder occurs at a
non universal value of the Luttinger parameter, which depends on the
disorder strength. This result places the transition in a universality class
different from the weak disorder transition first analyzed by Giamarchi and Schulz
[Europhys. Lett. {\bf 3}, 1287 (1987)].
While the details of the disorder potential are unimportant at the critical point,
the type of disorder does influence the properties of the
insulating phases. We find three classes of insulators which arise for different
classes of disorder potential. For disorder only
in the charging energies and Josephson coupling constants, at integer filling
we find an incompressible but gapless Mott glass phase.
If both integer and half integer filling factors are
allowed then the corresponding phase is a random singlet insulator,
which has a divergent compressibility. Finally in a generic disorder potential
the insulator is a Bose glass with a finite compressibility.

\end{abstract}

\maketitle

\section{Introduction}

%The superfluid insulator transition (SIT) has played a central role as
%a case study of symmetry breaking quantum and
%classical phase transitions \cite{sachdevbook}.
Superfluid-insulator transitions occur in a variety of
experimental systems, ranging from
low-temperature Helium through Josephson arrays
to ultra-cold atomic systems.
The simplest paradigm of such a transition is the rather well understood
Mott transition of interacting bosons on a perfect lattice commensurate with
the boson density. \cite{SachdevBook,giamarchi}. The theoretical picture is far less clear
in disordered systems, which occur in a wide variety of experiments:
Helium in Vycor, superconductor-metal
and superconductor-insulator transitions in nanowires and thin films \cite{Reppy1,chan-reppy,chan2,Lau,Bezryadin1,Bezryadin2,Kapi1,Kapi2,Kapi3,SteinerKapi,Yoon,Shahar1,Shahar2}.
Recently, disordered systems were also
realized using ultracold atoms \cite{zoller,atomchip-localization,Inguscio,Aspect,Bongs,DeMarco}.
Furthermore, this topic was brought back into the limelight with recent experiments
in solid Helium-4, which show the appearance of a superfluid fraction~\cite{Chan}. One
suggested explanation for this phenomenon is Helium turning superfluid
in structural defects of the surrounding solid \cite{Prokof'ev_ss}.
%The problem of disordered SIT was also address numerically,
%and theoretically, by numerous authors \cite{monien,Scalettar,Bray-AliMoore,WeichmanNew} (add citations).

Of particular interest is the superfluid-insulator transition
in disordered one-dimensional systems. Even without the disorder
the superfluid phase in one dimension is more subtle than in high dimensions.
In particular it does not exhibit true long range order. Nevertheless the uniform superfluid
admits a simple description in terms of a universal harmonic theory, or Luttinger liquid.
In the opposite limit of a disordered potential but no interactions, particles are always localized.
One might naively guess that there is no superfluid phase in the presence of disorder since interaction
alone or disorder alone both have a localizing effect on the bosons. This however does not seem to be the case.
%Since either disorder or interaction alone each have a localizing effect on the Bose system
%it might be naively expected that

The simplest way to see this is to introduce disorder as a perturbation to the
interacting superfluid within the Luttinger liquid description. This was done
by Giamarchi and Schulz in Refs. ~[\onlinecite{Giamarchi-Schulz1987,Giamarchi-Schulz1988}].
The main result of this approach is to describe a phase transition between an essentially uniform superfluid, in which the disorder is irrelevant, into a localized phase.
The natural tuning parameter of the transition is the interaction constant and it occurs at a {\em universal value} of the Luttinger parameter, independent of the strength of the disorder.

The above approach suffers from two main limitations. First, because it is perturbative in the disorder strength localization is signaled by a runaway RG flow.  Therefore the approach does
not allow for a detailed theory of the insulating phase. Second, the natural regime for the phase transition in this analysis is that of
strong interactions and a nearly uniform superfluid, which is not always the case in systems of interest.
For example atom-chip traps, in which ultracold atoms seem to undergo a localization
transition\cite{atomchip-localization}, are in precisely the opposite regime. The bosons are weakly interacting, while the potential they feel is highly disordered\cite{DW-Wang}. It is not clear whether the analysis of Giamarchi and Schulz provides a valid description of the transition in such a system.
%It is not clear that the same fixed-point describes the transition in the opposite regime of weakly interacting bosons subject to strong disorder. This is the
%regime naturally realized

Different approaches have been used to specifically describe the insulating phases of bosons
and suggested several possibilities depending on the nature of the system.
In the most generic disordered potential, Ref. [\onlinecite{FWGF}] argued for the
formation of a Bose glass phase characterized by a finite compressibility and
diverging local superfluid susceptibility. In the presence of a commensurate lattice
Refs. ~[\onlinecite{LeDoussalGiamarchi1,LeDoussalGiamarchi2}] predicted the existence of a Mott-glass phase,
an incompressible yet gapless insulator.

In recent work we introduced a unified approach to treat both the phase transition at strong disorder
as well as the properties of the insulating phases\cite{AKPR,AKPR2}. For this purpose we
employed a real space renormalization group (RSRG) technique \cite{MaDas1979,DSF94,DSF95, singh}.
%A similar approach was also used very recently to describe a dissipative superconducting-metal
%transition in disordered wires~\cite{VojtaHoyos1,VojtaHoyos2}.
We found that the superfluid insulator transition at strong disorder
is insensitive to the type of disorder introduced into the system. It is always Kosterlitz-Thouless like
in the following sense: characteristic time scales and length scales both diverge at the transition
as $\exp(1/\sqrt{\a-\a_c})$, where $\a$ is the tuning parameter.  The nature of the disordered
superfluid phase is also universal. It is described by an effective harmonic chain with random
Josephson couplings drawn from universal distributions generated as fixed points of the RSRG flow.
These distributions were recently used to compute the localization behavior of density waves\cite{GRC}.

The symmetry properties of the disorder, while not important in the superfluid phase or the transition, are
crucial for determining the nature of the insulating phases. Using the RSRG approach
we confirmed the formation of a Bose glass phase
for generic disorder and a Mott glass for a commensurate lattice with off-diagonal disorder.
The latter phase was also seen in recent numerical simulations~\cite{prokofiev, haas}.
In addition we found a novel glassy phase, which we termed a random-singlet glass,
in a system with particle hole symmetry.
This phase is characterized by a divergence of both compressibility and superfluid susceptibility.
Nevertheless it is still insulating, with conductance dropping as $\exp(-\sqrt{L})$ with
length. This phase is analogous to the random-singlet phase found in the spin-$\half$ X-Y chain~\cite{DSF94}.

The purpose of the present paper is twofold. First, we provide the detailed analysis of
chains with generic diagonal disorder, leading to the results
of Ref. [\onlinecite{AKPR2}]. Second we extend the analysis and compute the value of the Luttinger parameter
at the phase transition within the RSRG method. We find that, at strong disorder, the transition occurs at a non universal value of the Luttinger parameter that depends of the strength of disorder. This is contrary to the
perturbative analysis of Refs. [\onlinecite{Giamarchi-Schulz1987,Giamarchi-Schulz1988}].

%We were not able to fully resolve the origin of discrepancy between our results and those cited above. At strong disorder we show that the critical point corresponding to SIT corresponds to a broad distribution of the tunneling amplitudes between the sites. Thus disorder can not be treated perturbatively and the replica approach, which relies on weak Gaussian disorder and which predicts universal jump in Luttinger liquid parameter, is not applicable. We support our results using toy model examples, where we illustrate that the broad distribution of couplings indeed naturally arise in 1D systems.

The structure of the paper is as follows. In section \ref{sec:model}
we define the model we study and discuss its relevance to actual physical systems.
We give a detailed derivation of the RSRG flow equations for the special case of
particle-hole symmetric disorder in section \ref{phs} and for generic disorder
in section \ref{genericsec}. We give a detailed account of the numerical as well as
the approximate analytical solutions of the flow equations. Then in section
\ref{SFLutt} we solve for the value of the Luttinger parameter at the transition.
Finally we conclude with a summary of the results and a discussion of their possible
experimental implications.

\section{Model}\label{sec:model}

Our starting point for the theoretical analysis is the quantum rotor Hamiltonian
\be
\H=\sum_j U_j \left(\hat{n}_j-\overline{n}_j\r)^2 -\sum_j J_j
\cos\l(\f_{j+1}-\f_j\r). \label{model}
\ee
This model describes an effective Josephson junction array with random Josephson
coupling $J_i$ and charging energies $U_i$. In addition there is a random offset charge
${\bar n}_i$ to each grain, which is tantamount to a random gate voltage.
Although the model can be visualized as a Josephson junction array, it actually provides an
effective description valid for a wide variety of systems that undergo a superfluid to insulator transition.
In bosonic systems in particular, such transitions are usually driven by quantum phase fluctuations.
The hamiltonian (\ref{model}) should then be thought of as a low energy
effective theory one obtains after integrating out the gapped amplitude fluctuations. The remaining
degrees of freedom relevant to the transition are the quantum rotors.

One concrete example of how such a model is naturally generated at low energies is
provided by a system of ultracold atoms in an atom-chip trap.
In this system the disordered potential
is induced by corrugation in the wire that generates the trapping magnetic field\cite{DW-Wang}.
With increasing corrugation, the atoms concentrate in small puddles at minima of the potential.
Neighboring puddles are connected with each other by a random Josephson coupling which depends
on the potential barrier between them. The result is exactly the random Josephson array
defined in Eq. (\ref{model}).

Another possible physical realization of the model (\ref{model}) is
a disordered superconducting nano-wire. Here the issue is more subtle because there may be
gapless Fermionic degrees of freedom that generate dissipation. Indeed Refs.
\cite{VojtaHoyos1,VojtaHoyos2},
applied the RSRG to such wires starting from a Hertz-Millis\cite{Hertz,Millis} dissipative action, with a dissipation
term $|\w|\psi^*\psi$. An alternative approach is to use phase only models, which describe resistively shunted Josephson junction arrays\cite{RDOF2}. This naturally leads to a dissipation term of the form $q^2|\w|\phi^*\phi$, which
does not affect global superconductor insulator transitions. Such models combined with strong disorder
may also be described by the present analysis in parts of their phase diagram.

\section{Particle-hole symmetric chemical potential disorder \label{phs}}

Of all the random coupling constants in the model (\ref{model}), the random offset charge (or local chemical potential) seems to be the hardest to incorporate in an
RG treatment. Since the offsets simply add up to give the offset of a block
of sites, it seems clear that this disorder will just grow as the square root
of the scale of the real space RG making it hard to track.  However this difficulty
turns out to be  largely
superficial and can be easily overcome in the analysis.
To make the discussion more transparent we start from the case where only integer and half-integer offset charges, $\nbar_i$, are
allowed on each site. This condition maintains particle-hole
symmetry, and therefore still does not correspond to the generic
case. Nevertheless, this restriction allows a relatively simple RG
analysis which affords important analytic and numerical insights
into the possible phases and the phase transitions. In the next section we generalize our treatment to the case of generic disorder.

We note that despite the restrictive condition, allowing  only $\nbar_j=0,1/2$, this
type of disorder may actually be a reasonable approximation for chains of
superconducting grains with pairing gap much larger than the charging energy.
Under these conditions we can assume that the electrons on the grains are always
paired and we can take $e^*=2e$ as the unit of the bosonic charges. On the other hand the positive background charge, is a random number that could be even or odd in units of $e$ and consequently either integer or half integer in units of the boson charge $e^*$. Allowing for charged impurities on the substrate or unscreened coulomb interactions between different grains, would of course lead violation of the restrictions on the off set charges.

\subsection{Particle-hole symmetric quantum rotor model}

The essence of the renormalization group transformations either in
real or momentum space is the gradual coarse graining of the system.
In this section we extend the decimation scheme of
Ref.~\cite{AKPR} to the Hamiltonian in Eq. (\ref{model}) for the case that $\overline{n}_j$ can take the values of $0$ and $1/2$ randomly. The last condition ensures the particle-hole symmetry in the problem: the Hamiltonian
does not change under the transformation $n\to 1-n$. These two
values of $\overline{n}_j$ represent the two possible extremes which
drive the physics of the Bose-Glass~\cite{FWGF}. Sites with
$\on_j=0$ have a well defined Coulomb blockade with charging energy
$U_j$. Sites with $\on_j=1/2$, on the other hand, have no Coulomb
blockade. With no further interactions, these sites yield both
infinite compressibility and infinite superfluid susceptibility due
to the number fluctuations costing no energy. The
Hamiltonian~(\ref{model}) is characterized by the distribution of
hoppings $J_j$ and charging energies $U_j$, and of the proportion of
sites with $\on=1/2$. We will refer to the latter sites as
half-integer sites or 'half-sites'.

\subsection{Extended real-space renormalization group \label{ERG}}

Let us now construct the extended decimation scheme for the
model~(\ref{model}). Following Refs.~[\onlinecite{DSF94, DSF95, MaDas1979,AKPR}] we construct an RG scheme that eliminates
iteratively large energy scales from the Hamiltonian. Two sites
connected by the strongest bond will be converted to a
phase-coherent cluster. Similarly, in sites with strong charging
energy $U$ we eliminate all the excited states. However, the result
of this elimination will be different for integer and half-integer
sites. Let us now discuss these steps in detail.

We denote the largest energy scale in the Hamiltonian (\ref{model})
$\Omega=\max\{J_i,U_i\}$. In each step in the RG we eliminate the
strongest coupling from the Hamiltonian, and hence successively
reduce $\Omega$. If the strongest coupling is the charging energy of
site $i$, $U_i$,  we eliminate all the excited states of this site.
For integer sites with $\on_i=0$, we minimize the charging energy by
setting $n_i=0$ and include the coupling of this site to the rest of
the chain perturbatively. As in Ref.~[\onlinecite{AKPR}], the second
order perturbation theory leads to a new coupling between the new
nearest neighbors $i-1$ and $i+1$:
\be
\tilde{J}_{i-1,\,i+1}=J_{i-1}J_{i}/\W. \label{jrecurs}
\ee
On the other hand, if $\on_i=1/2$ we reduce site's $i$ Hilbert space
to the states $n_i=0,\,1$. The hoppings connecting site $i$ to its
neighbors are still active, and to the first approximation are not
affected by the elimination of the high energy states. The
decimation step for $\on_i=1/2$ produces a new kind of site, a {\it
doublet site}, only capable of having $n_i=0$ or $1$. Let us denote
the fraction of doublet sites as $\s$, the fraction of integer sites
as $\q$, and the fraction of half-sites as $\p$. Note that these
fractions add up to unity $\p+\q+\s=1$.

When the strongest coupling in the chain is the bond $J_i$, unless
{\it both} sites $i$ and $i+1$ are already-decimated doublet sites, a
phase-coherent cluster forms. Since charging energy is the inverse of
capacitance, the effective $U_{i,\,i+1}$ of the new cluster will be:
\be
\frac{1}{\tilde{U}_i}=\frac{1}{U_i}+\frac{1}{U_{i+1}}
\label{urecurs}
\ee
For a doublet site, $U_{i}$ is set to $\infty$. It is easy to see that the filling
factor $\overline{n}$ is an additive quantity:
\be
\tilde{\on}_{i,\,i+1}=(\on_i+\on_{i+1})\mod 1.
\label{nrecur}
\ee
Therefore two half-sites or two integer-sites form an
integer-cluster. An integer site and a half-site form a
half-cluster. Similarly, a doublet site and an integer site
form a half-cluster, and a doublet and  half-site form an
integer cluster.

It is important to note here that the above decimation step {\it does
  not} assume long range order; it states that phase
  fluctuations within the newly-formed cluster are harmonic, and
  therefore the cluster can not be broken due to phase-slips. These
  harmonic fluctuations are crucial for the understanding of the
  properties of the superfluid phase, as explained in
  Sec. \ref{SFLutt}. Nevertheless these phase fluctuations
can be neglected for the purpose of the RG flow, and they do not
  change the critical properties of the model \cite{AKPR}.

A qualitatively new decimation step, which goes beyond
Ref.~[\onlinecite{AKPR}] occurs when the strong bond $J_i$ connects
two doublet sites. In this case the two sites form a unique
non-degenerate ground state:
\be
\ket{\psi_{i,\,i+1}}={\ket{n_i=0,n_{i+1}=1}+\ket{n_i=1,n_{i+1}=0}\over
\sqrt {2}},
\ee
which has energy $-J_i/2$. The second order perturbation theory
leads to an effective hopping between sites $i-1$ and $i+2$:
\be
\tilde{J}_{i-1,\,i+1}=J_{i-1}J_{i+1}/J_{i}=J_{i-1}J_{i+1}/\Omega. \label{jrs}
\ee
Since each doublet site can be thought of as a spin-1/2 degree of
freedom, the elimination of $J_i$ consists of the formation of a
singlet. Hence we recover the Ma-Dasgupta RG transformation~\cite{MaDas1979,DSF94}. Note that formally Eqs.~(\ref{jrs}) and (\ref{jrecurs}) are identical.

\subsection{Flow equations}

Next, we describe the flow equations implied by the above decimation steps.
As in Ref.~[\onlinecite{DSF94,DSF95}], we parametrize the cutoff energy
scale with the variable $\G\equiv\log(\W_0/\W)$, where $\W_0$ is the
initial cutoff. Also, we define the dimensionless couplings
$\z_i=\W/U_i-1$ which are characterized by probability distributions $f_q(\z,\G)$ for integer-sites, and $f_p(\z,\G)$ for half-sites. In principle these distributions can be different, but one can show
that their difference is irrelevant in the RG sense, and therefore does
not affect any of our conclusions. For simplicity, we assume from
the beginning that $f_q=f_p\equiv f$. We also define
$\b_i=\log(\W/J_i)$ as the logarithmic bond variable, with
distribution $g(\b,\G)$. Note that by construction $\b_i$ and $\z_i$ have nonzero probability distribution in the interval $[0,\infty)$.

Renormalization group steps gradually
decrease the number of remaining sites in the chain ($N(\Gamma)$).
Thus decimation of the integer site with large charging gap $U$
reduces $N$ by one while a similar decimation of a half-integer site
simply converts it to the doublet site. Also, decimation of a strong
link and joining two sites into a cluster reduces the number of
active sites by one unless the link connects two doublets. In the
latter case the number of remaining sites is reduced by two. Thus
the flow of $N$ is given by
\be
d N(\Gamma)=-\left[g_0(\Gamma)(1+s^2)+\q f_0(\Gamma)\right]
N(\Gamma)d\Gamma,
\label{rhoflow}
\ee
where $f_0(\Gamma)\equiv f(0,\Gamma)$ and $g_0(\Gamma)\equiv
g(0,\Gamma)$. From Eq.~(\ref{rhoflow}) and the above RG conditions,
we obtain the flows of the fractions $\p,\,\q$, and $\s$:
\beq
&&\frac{d\s}{d\G}=-g_0\s\l(1-\s^2\r)+ f_0 (\p+\q\,\s)\nonumber\\
&&\frac{d\p}{d\G}=-g_0\l[\p\l(1-\s^2\r)-2\q\l(1-\q\r)\r]-f_0(\p-\p\,\q)
\label{pqsflow}\\
&&\frac{d\q}{d\G}=-g_0\l[\s^2-1+3\q-2\q^2-\q\s^2\r]-f_0(\q-\q^2).\nonumber
\eeq
It is easy to check that $d\s/d\G+d\p/d\G+d\q/d\G=0$, provided that $\p+\q+\s=1$.

The RG conditions also lead to master equations for the distributions:
\begin{widetext}
\beq
&&{\partial f(\z)\over \partial \G}= (1+\z){\partial
f(\z)\over\partial \z}+(1-\s)g_0\int\int d\z_1 d\z_2 f(\z_1)
f(\z_2)\delta(\z_1+\z_2+1-\z)-f(\z)g_0(1-\s)+f(\z)(f_0+1)\nonumber\\
&&{\partial g(\beta)\over \partial \G}= {\partial
g(\beta)\over\partial \beta}+\l(\s^2 g_0+\q f_0\r)\int\int d\beta_1
d\beta_2\, g(\beta_1) g(\beta_2)\delta(\beta_1+\beta_2-\beta)
+g(\beta)g_0(1-\s^2)-\q g(\beta) f_0,
\label{master}
\eeq
\end{widetext}

Even though these equations look quite complicated, the meaning of
the each term is straightforward. For example, the second term in
the first of these equations corresponds to renormalization of the
capacitance of the cluster following the decimation of the link. The
multiplier $1-s$ reflects the fact that the renormalization takes
place only if the link does not connect two singlets.

The equations (\ref{pqsflow}) and (\ref{master}) can be
significantly simplified noting that $\p=\q=(1-\s)/2$ is their
solution for arbitrary functions $f_0(\G)$ and $g_0(\G)$. It is
easy to check that $\p=\q$ is in fact an attractive solution. Indeed
Eqs.~(\ref{pqsflow}) give:
\be
\frac{1}{\q-\p}\frac{d(\q-\p)}{d\G}=-f_0(1-\q)-g_0\l(2(1-\q)+\s(1-\s)\r).
\label{p-q}
\ee
Unless $q=1$, the RHS of Eq.~(\ref{p-q}) is always negative, which
means that the line $\p=\q$ is an attractor. Physically one can
understand this result as follows: The integer versus half integer
filling of a cluster is determined by the parity of the total number
of half-integer decimated sites. As clusters grow in size under RG
due to coarse-graining, the number of such sites becomes large, and
thus even and odd parities occur with the same probability. The case
of commensurate disorder,
$q=1$, which was analyzed in Ref.~[\onlinecite{AKPR}] is an
exception, since it corresponds to the strictly zero fraction of
half-sites where the clusters always remain even.

The other important observation is that in the weakly interacting
regime $f_0\ll 1$ one can use a simple exponential ansatz to solve
Eqs.~(\ref{master}):
\be
f(\z)=\p f_0 e^{-f_0\z}\quad g(\beta)=g_0e^{-g_0\beta}
\label{scaling}
\ee
As we will see below (see also Ref.~[\onlinecite{AKPR}]), the
universal properties of the superfluid-insulator transition are
determined by the noninteracting fixed point with vanishing $f_0$,
where the ansatz is well justified. According to our numerical
simulations, these exponential scaling functions are attractors of
the flow equations, and they describe very well the distribution of
$\zeta$ and $\beta$ even when $f_0$ is not very small (see
discussion below). Substituting the ansatz (\ref{scaling}) and
$\p=\q=(1-\s)/2$ into Eqs.~(\ref{pqsflow}) and (\ref{master}) we
find:
\beq
\frac{d f_0}{d\G}&=&f_0\l[1-g_0(1-\s)(1+f_0)\r],\nonumber\\
\frac{d g_0}{d\G}&=&-{g_0\over 2}\l[(1-\s)f_0+2\s^2 g_0\r],\label{redeq}\\
\frac{d\s}{d\G}&=&{f_0\over 2}(1-\s^2)-g_0\s(1-\s).\nonumber
\eeq
This system has two fixed points for $\s$: $s=0$ and $s=1$. The first fixed point $s=0,\,p=q=1/2$ as we will see below describes the superfluid phase, while the second one: $s=1,\,p=q=0$ corresponds to the random-singlet glass insulator.

\subsection{SF fixed point \label{FPscale}}

Let us first address the superfluid fixed point - $s=0$ (no doublet sites).
Note that from the last
equation of the system (\ref{redeq}), this fixed point is
stable only when $f_0$ is small. Since  $f_0$ flows either to zero or to
infinity, the fixed point $\s=0$ can is stable when $f_0\to
0$. Then the linearized flow equations to first order in $s$ and
$f_0$ reduce to:
\beq
\frac{df_0}{d\G}&=&f_0 (1-g_0)\label{SFeqa}\\
\frac{dg_0}{d\G}&=&-\frac{1}{2}f_0.
\label{SFeq}
\eeq
Remarkably, apart from a factor of $1/2$ in the second equation,
these are the same flow equations as obtained near the SF fixed
point of the random Bose-Hubbard model with no half-sites (i.e.
$\q=1$), as in Eq.~(6) of Ref.~\cite{AKPR}. The extra factor
of $1/2$ can be absorbed in a redefinition of $f_0$. This factor
appears because half of the sites in the lattice have half-integer filling and hence their interactions are ineffective in suppressing the superfluidity and thus do not renormalize $g_0$ which is related to SF stiffness (see Sec.~\ref{SFLutt}).

Equations (\ref{SFeqa}) and (\ref{SFeq}) are easily solved to give
\be
f_0(\G)=A+2 g_0(\G)-2\ln g_0(\G)-2=A+\left[1-g_0(\Gamma)\right]^2.
\label{crit_flow}
\ee
where $A$ is a tuning parameter that controls the flow. When $A<0$,
the flows terminate on the non-interacting fixed line
$f_0=0,\,g_0>1$.

The point $A=0$ lies on the critical manifold, which
terminates in a K-T fixed point (note that if we use $v=\sqrt{f_0}$ we
obtain the standard Kosterlitz-Thouless flow equations) at
$g_0=1,\,f_0=0$.

At the critical manifold we can solve exatly for the flow of
$g_0$ and $f_0$. Using the parametrization $g_0=1+\delta g$. $f_0$ and $\delta
g$ are determined by the differential equations:
\be
\ba{c}
\der{\delta g}{\Gamma}=-\frac{1}{2}f_0 \vspace{2mm}\\
\der{f_0}{\Gamma}=-f_0 \delta g\vspace{2mm}\\
\ea
\ee
By dividing the two equations we have:
\be
d(\delta g)^2=df_0
\ee
and:
\be
\der{\delta g}{\Gamma}=-\frac{1}{2}\delta g^2.
\ee
This gives:
\be
\ba{cc}
\delta g=\frac{2}{\G}+\frac{2}{\G^2} \hspace{5mm}& \hspace{5mm} f_0=\frac{4}{\G^2}.
\ea
\ee
Note that the solution of the flow equations in the case of only
integer sites (q=1) yields the same results, only with $f_0\rightarrow 2/\G^2$.

\begin{figure}[ht]
\includegraphics[width=8cm]{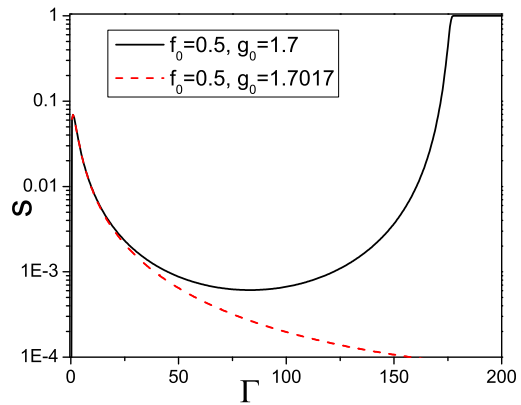}
\caption{RG flow of the fraction of the singlet
sites $s$ according to Eqs.~(\ref{redeq}) near the critical point.
The solid line corresponds to the insulating regime, while the
dashed line does to the superfluid phase. In both cases we assumed
there are no singlet sites at the onset, i.e. $s(\G=0)=0$.}
\label{fig_sflow}
\end{figure}

When $A>0$, the parameters $f_0$ and $g_0$ flow past the fixed point,
where $f_0$ begins to increase. The increase of $f_0$
entails a flow away from the $s=0$ fixed point, towards the $s=1$
fixed point. In Fig.~\ref{fig_sflow} we show the examples of flows
of $\s$ near the critical point both in the superfluid and in the
insulating regimes.

We now want to stress a very important point. It appears that in the
superfluid regime the system flows to the classical fixed
line, where there is no charging term. However, as we hinted above (see Sec. \ref{ERG}), this statement should be understood with special care. Each time the RG scheme merges
two sites into a single superfluid cluster it neglects the
harmonic Josephson plasmon between the two sites, although its energy is well below the RG cutoff at this step. These internal excitations
do not influence the progression of the RG flow. However, they become the elementary phonon
excitations of the single superfluid cluster that evolves to be the RG fixed point.
Therefore the fact that the charging term becomes irrelevant simply means that one should keep {\em only} these harmonic phonons. In other words, one can ignore vortices or phase-slips, which destroy the superfluid phase if they proliferate. Indeed there is a direct connection between onsite interactions and phase slips. As we show next, at strong disorder, such sites are responsible for renormalization of the superfluid stiffness $\rho_s$ playing the role of phase slips. The fact that $f_0$ flows to zero implies that such events renormalizing $\rho_s$ become unimportant
and one can use a noninteracting quadratic description. This issue
will be discussed thoroughly in Sec. \ref{SFLutt}. Technically the fact that interactions are irrelevant in our description comes from the fact that we are working in the grand-canonical ensemble. While in the insulating regime there is no difference between excitation energies in canonical and grand-canonical ensembles, in the superfluid regime there is a significant difference. Thus in canonical ensemble the lowest energy excitation corresponds to a phase twist or phonon while in the grand-canonical ensemble the lowest energy corresponds to the addition of an extra particle, which costs much less energy than the phase twist. So the fact that in our scheme the interactions are irrelevant in the SF phase should be understood only in this grand-canonical sense.

\subsection{Insulating fixed point - random-singlet glass}

The $s=1$ fixed point corresponds to the insulating phase. Indeed
one can check that in this limit $g_0(\G)\to 0$ and
$f_0(\G)\to\infty$. However, this insulator is not the Mott-glass
that describes the case of integer only sites considered in
Ref.~[\onlinecite{AKPR}]. At $s=1$, all the sites remaining in the
system are doubly degenerate. These sites can be thought of as
spin-1/2 degrees of freedom, with $\ket{\uparrow}=\ket{\on+1/2}$ and
$\ket{\downarrow}=\ket{\on-1/2}$. Without hopping, the ground state
obviously has a huge degeneracy. However, the residual hopping lifts
this degeneracy. In the spin language, the hopping terms correspond
to usual $xy$ spin-spin interaction:
\be
J_i\cos\l(\phi_{i+1}-\phi_i\r) \to J_i\l(\sigma^x_i\sigma^x_{i+1}+\sigma^y_i\sigma^y_{i+1}\r)
\ee
Thus we arrive at a spin-1/2 system with random $xy$ couplings. The
ground state of this system is known to be the random-singlet
phase~\cite{DSF94}. A strong bond between sites $i$ and $i+1$
delocalizes a boson between two sites, and creates a cluster that
has a charge gap $J_i$:
$\ket{\psi_{i,\,i+1}}=\ket{1_i}\ket{0_{i+1}}+\ket{0_i}\ket{1_{i+1}}$.
Quantum fluctuations produce an effective coupling between sites
$i-1$ and $i+2$ as in Eq.~(\ref{jrecurs}). The typical length over
which the singlets form, $\ell$, scales as $\G^2$. Alternatively,
one can say that the gaps of each singlet-cluster is
$\Delta=\W_0\exp(-\sqrt{\ell})$.

The flow equations can be linearized near $s=1$, and we obtain:
\beq
&&\frac{df_0}{d\G}=f_0\nonumber\\
&&\frac{dg_0}{d\G}=-g_0^2\label{in_flow}\\
&&\frac{d\s}{d\G}=(1-\s)f_0\nonumber
\eeq
This system implies that $f_0$ diverges as $f_0\sim e^{\G}$. This divergence implies that interaction in the remaining non-singlet sites is narrowly distributed near the maximum energy scale $\Omega$. Following from that scaling of $f_0$, $\s$ converges to $1$ extremely fast: $\s=1-\delta \exp\l(-e^{\G}\r)$. And finally $g_0$, which corresponds to the average of $\ln(J_j)$, follows the random singlet scaling, and flows slowly to zero as: $g_0=\frac{1}{\G}$.

The random-singlet glass is an insulator with the superfluid
stiffness of a chain of length $L$ scaling as \cite{DSF98}
\be
\rho_s\sim e^{-C\sqrt{L}}
\ee
with $C$ being a nonuniversal constant. This behavior of $\rho_s$
immediately follows from Eqs.~(\ref{rhoflow}) and (\ref{in_flow}),
see also Ref.~[\onlinecite{DSF94}]. At the same time this insulator
is gapless, with the gap also decaying exponentially with $\sqrt{L}$
but with the coefficient $C/2$. Unlike the Mott glass phase or a
Bose glass phase, which we will discuss below, the random singlet
insulator is characterized by a diverging density of states at zero
energy and hence by a divergent compressibility ($\kappa$) and
superfluid susceptibility $\chi_s$. The former $\kappa=dn/d\mu$, in
the spin language is the response to a field $\sigma^z \delta\mu$.
Similarly, $\chi_s$ defined as the response to the perturbation
$\delta(a+\a^\dagger)$, in the spin language, corresponds to the
perturbation $\sigma^x \delta\psi$. Since the random singlet ground
state has SU(2) symmetry, the two responses have the same form, and
diverge as:
\be
\kappa,\chi_s\sim\frac{1}{\delta\mu\log^3(\W_0/\delta\mu)},\quad
\lambda\sim\frac{1}{\delta\Delta\log^3(\W_0/\delta\Delta)}.
\ee
As the slow decay of $g_0$ with $\G$ suggests, indeed the
random-singlet glass has more superfluid features than the
Mott-glass. Both have a vanishing gap, but the Mott-Glass has vanishing compressibility, and its superfluid-susceptibility is
only finite.

\subsection{Numerical RSRG for the p-h symmetric state}

\begin{figure}
\includegraphics[width=8cm]{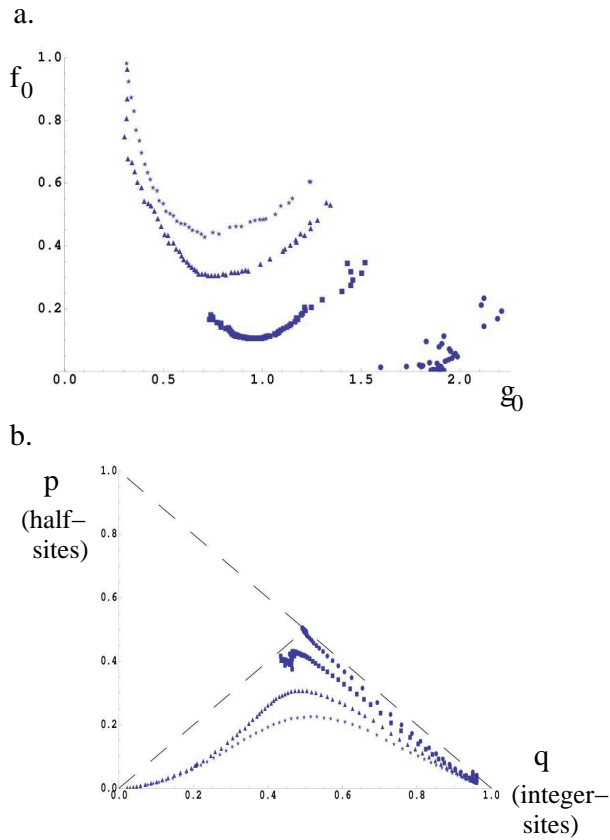}
\caption{RG flows of various realizations of disorder in the  (a)
  $f_0$ - $g_0$ plane, and (b) p-q plane,  for chains with initial
distributions characterized by: $(\delta,p_0)=(-0.1,0.05)$ (circles),
$(-0.05,0.04)$ (squares), $(-0.02,0.04)$ (triangles), $(0,0.04)$ (stars). Here $p_0$ is the initial
fraction of the half integer p-sites. \label{fgfig1}}
\end{figure}

In order to corroborate the analytical results for the RSRG, we
carried out the RG flow numerically without any simplifying
assumptions. The numerics, by and large, backs the analytical
results. In Fig.~\ref{fgfig1} several flow traces are given in the
$f_0$ vs. $g_0$ parameter space, and in the
$p-q$ plane. The initial distributions used consist of box
distributions in the range $0.2+\delta<U<1.2+\delta$ for the charging
energy, and $0.2-\delta<J<1.2-\delta$ for the nearest-neighbor
tunneling.

%-------------------- full model

\section{Phase diagram of the B-H model with generic disorder \label{genericsec}}

When considering experimentally realizable models, we must also consider
randomness in the chemical potential, or random offset charges. In
particular $\on$ could have values anywhere between $-1/2$ and
$1/2$. Typically, this type of disorder is very relevant. Indeed if
we join two sites $1$ and $2$ together into a cluster then the new
value of $\overline n_{12}$ becomes a sum of $\overline n_1$ and
$\overline n_2$ modulo one (so that the result also belongs to the
(-1/2, 1/2] interval).

To address this problem it is worthwhile review a few important insights gained
from our analysis of the particle hole symmetric model.
In that case the low energy behavior was dominated by the line $p=q$ where the
number of half-sites with $\on=1/2$ was the same as the number of
integer-sites, with $\on=0$. At the same time we saw that the
universal properties of the SF-INS transition with $p=q$ and $p=0,\;
q=1$ were identical up to a factor of one half in Eq. (\ref{SFeq}),
which is absent in the integer filling model.

We thus can anticipate that the fixed point governing the
SF-insulator transition has a uniform distribution of $\on$ when we
remove the particle-hole symmetry restrictions. But also, in analogy
with the half integer case, we can expect that the fixed point
describing the SF-INS transition remains intact. On the other
hand, again having the p-h symmetric case in mind, we expect that
the distribution of $\on$ at the critical point strongly affects the
properties of the insulating phase. As it turns out, the diagonal
disorder plays the role of a 'dangerously irrelevant variable' (as
$\s$ is in the analysis above - irrelevant in the SF side of the
transition but strongly relevant in the insulator side). The
diagonal disorder does not change the nature of the critical and
crossover behavior, but it determines to which insulating phase the
system will flow. In Ref.~\cite{AKPR} the p-h symmetric model
with only integer fillings had a Mott-glass insulating phase. By
allowing also sites with charging degeneracy ($\on=1/2$) but
preserving the p-h symmetry, the system in its insulating phase is a
random-singlet glass. When we remove the p-h symmetry, we expect
that the insulator becomes a Bose-glass: gapless, compressible state
with a diverging susceptibility to SF fluctuations.

Before going into more detailed analysis, which confirms the
above assertions, we would like to comment on the similarities with
the perturbative RG approach of Giamarchi and Schulz~\cite{Giamarchi-Schulz1988}. In
particular they derived the following flow equations near the
transition between superfluid and localized phases:
\beq
&&{d\mathcal D\over d\Gamma}={9\over 2}\left(K^{-1}-{2\over
3}\right)\mathcal D,
\label{gs1}\\
&&{d(K^{-1})\over d\Gamma}={1\over 2} \mathcal D\label{gs2}
\eeq
where $K$ is the Luttinger parameter, $\sqrt{\rho_s \kappa}$ and $\mathcal D$ is proportional to the variance of the disorder in the chemical
potential: $\overline{\mu(x)\mu(x^\prime)}\propto \mathcal
D\delta(x-x^\prime)$. We point out that in
Ref.~[\onlinecite{Giamarchi-Schulz1988}] the Eqs. (\ref{gs1}-\ref{gs2}) are written in terms of the inverse Luttinger parameter being $K$, the inverse of the common convention which we use (see e. g. Ref.~\cite{prokofiev}). Note that there is a direct analogy between Eqs.~(\ref{SFeqa}), (\ref{SFeq}) and
Eqs.~(\ref{gs1}) and (\ref{gs2}) if one identifies $K$ with $g_0$ and $\mathcal D$ with $f_0$.

Interestingly in Ref.~[\onlinecite{FWGF}] it was argued that the
SF-INS transition described by Eqs. (\ref{gs1}) and (\ref{gs2}) does
not belong to the KT universality class because of the first power
of $\mathcal D$ appearing in the second of these equations, as
opposed to the second power in the conventional case. This
difference according to the authors lead in particular to the
unconventional scaling of the correlation length with $K-K_c$ near
the critical point. However, this must be a misstatement since,
as we argued earlier, the substitution $\mathcal D=\sigma^2$
brings the flow equations to the conventional KT form.
This change of variables should not affect the scaling.
Moreover the flow equations in terms of $K$ and $\sigma$ are more
natural because $\sigma$ has dimensions of the external potential
and thus it (not $\mathcal D$) is analogous to the strength of the
periodic potential, which drives the transition in a nondisordered
case.

The similarity between the perturbative analysis of
Ref.~[\onlinecite{Giamarchi-Schulz1988}] and the one presented here goes even further.
A simple scaling argument shows that disorder in the chemical
potential is strongly relevant in both approaches. However in the
language of Ref.~[\onlinecite{Giamarchi-Schulz1988}] the strongly relevant part of the
disorder corresponds to the forward scattering, which can be
reabsorbed into the canonical smooth fluctuations of the density. It
is the backward scattering or phase-slips, which determine the fate
of the superfluid phase. By analogy with our approach we can argue
that even the smooth part of the disorder potential should
become strongly relevant in the
insulating regime. Thus it should play the role of a dangerously
irrelevant term just as the distribution of $\overline{n}$ does in our approach. Unfortunately
the pertubative RG approach becomes uncontrolled in the insulating
regime and this postulate cannot be reliably verified.

\subsection{RG scheme for the generic disorder B-H model}

We probe the observsations above by extending our RSRG analysis to treat
arbitrary disorder: $U_i$, $J_i$, and $\on_i$ all random in the model
in Eq. (\ref{model}).

First let us analyze the charging term while ignoring the hopping.
Each site has a charge gap given by:
\be
\Delta_i=\frac{1}{2}U_i\l(1-2|\on_i|\r)
\label{cgap}
\ee
where $-1/2<\on<1/2$.

As before, we treat the model iteratively, but this time, in each
step of the RG we find the largest energy scale:
\be
\W=\max_i\{J_i,\,\Delta_i\}
\label{ogap}
\ee
and eliminate it. If it is a gap, $\Delta_i$, then the site $i$
freezes into its lowest energy charging state. Quantum fluctuations
induce an effective hopping between sites $i+1$ and $i-1$:
\be
J_{i-1,\,i+1}=\frac{J_{i-1}J_i}{\W}{1\over (1+2|\on_i|)}.
\ee
The last multiplier in this expression is a non-universal prefactor,
which varies between $1$ and $1/2$. This prefactor does not affect
any universal features of the transition and we can safely set it to
unity. On the other hand, if $J_i$ is the largest energy scale, then
the sites $i$ and $i+1$ form a SF cluster with effective charging
energy given by Eq.~(\ref{urecurs}), and with a filling offset:
\be
\on_{i,\,i+1}=\on_i+\on_2
\ee
where the last equality is defined modulo adding or subtracting one,
so the the result always belong to the interval:
$\on_{i,\,i+1}\in(-1/2,1/2]$.

\subsection{Generic case flow equations}

The ensuing flow can be quantified using flow equations for the
distribution of logarithmic couplings, $\beta_i=\log\W/J_i$, and the
joint distribution $F(\zeta,\on)=f(\zeta,\on)\chi(\zeta-1+2|\on|)$
where $\zeta_i=\W/U$ and the Heaviside step function $\chi$ enforces
the constraint $\W/\Delta>1$ or equivalently $\zeta>1-2|\on|$.

The flow equations are given by:
\begin{widetext}
\beq
&&\frac{\partial g}{\partial\G}={\partial g\over \partial
\beta}+[f_1+g_0(1-f_s)]\,g\times
g\big|_{\beta=\beta_1+\beta_2}+g(g_0 f_s-f_1)\label{bgflow1}\\
&&\frac{\partial f}{\partial \G}=\zeta{\partial f\over
\partial \zeta}+g_0\, f\times
f\big|_{\zeta=\zeta_1+\zeta_2,\,\on=\on_1+\on_2}+f(1+f_1-g_0
f_s)\label{bgflow2},
\eeq
\end{widetext}
where
\be
f_1=\intt_{-0.5}^{0.5} d\on(1-2|\on|)f(1-2|\on|,\on)
\ee
is the density of sites with a large charging energy,
\be
f_s=\int d\zeta\int d\on\, \chi(\zeta-1+2|\on|)\,f\times
f\big|_{\zeta=\zeta_1+\zeta_2,\,\on=\on_1+\on_2}\,.
\label{fs}
\ee
Physically $1-f_s$ is the density of strong bonds connecting the
sites with large onsite interaction, which are close to half
filling. These sites form a cluster with $\Delta>\Omega$ and thus
have to be eliminated as a spin singlet. In the equations above
$\times$ implies the convolution over $\beta$ in $g\times g$ and
over both $\zeta$ and $\on$ in $f\times f$.

Although the equations look somewhat obstruse, near the critical
point they can be solved with the same scaling ansatz as before.
Indeed, since near the transition the interactions are negligible,
we can safely ignore $\delta\zeta$, which is of the order of one, in
the convolution (\ref{fs}). Also similarly to the particle-hole
symmetric case we can expect that near the critical point the
distribution of $\on$ is uniform and thus $f(\zeta, \on)$ does not
depend on $\on$. We then use our standard scaling ansazt:
\beq
&& g(\beta)=g_0\mathrm e^{-g_0\beta}, \label{ans1}\\
&& f(\zeta, \on)={f_0^2\over 1-\mathrm e^{-f_0}}\, \mathrm
e^{-f_0\zeta}\approx f_0\mathrm e^{-f_0\zeta}\label{ans2},
\eeq
where in the last equality we used $f_0\ll 1$. In the same
approximation of small $f_0$ we find that $f_1\approx f_0/2$.
Substituting the scaling ansatz into the flow equations
(\ref{bgflow1}) and (\ref{bgflow2}) and using $f_0\ll 1$ we
immediately recover that $f_0$ and $g_0$ obey Eqs.~(\ref{SFeqa}) and
(\ref{SFeq}). The latter automatically implies that the SF-IN
transition in the case of generic disorder belongs to the same
universality class as in the particle-hole symmetric case.

\begin{figure}
\includegraphics[width=8cm]{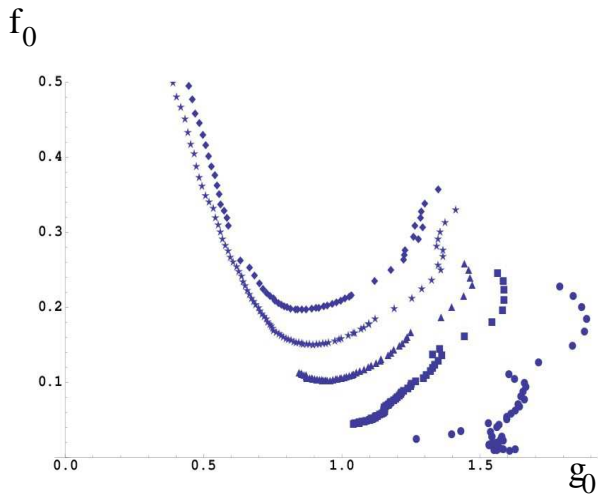}
\caption{Flows in the $f_0$ vs. $g_0$ plane. Initially, the interaction energy
  $U$ and the hopping $J$ are uniformally distributed in the range
  $0.2-\delta<J<1.2-\delta$, and $0.2+\delta<U<1.2+\delta$; the offset
  charge is also uniformally distributed between
  $-\Delta\on<\on<\Delta\on$. The values of $\delta$ and $\Delta\on$
  for the plots shown are:
$\delta=-0.05,\,\Delta\on=0.12$ which is in the SF phase (circles),
$\delta=-0.04,\,\Delta\on=0.08$ (squares),
$\delta=-0.03,\,\Delta\on=0.08$ (triangles),
$\delta=-0.02,\,\Delta\on=0.08$ (stars), and
  $\delta=0,\,\Delta\on=0.12$ (diamonds).
  \label{figrmu1}}
\end{figure}

We confirm these findings performing numerical analysis of the full
RG equations (\ref{bgflow1}) and (\ref{bgflow2}). We find a clear
signature of the K-T transition that is even more pronounced than
before. In Fig.~\ref{figrmu1} the flows in the f-g parameter space
are shown for three different initial conditions.

\subsection{Nature of the Bose-glass}

\begin{figure}
\includegraphics[width=8cm]{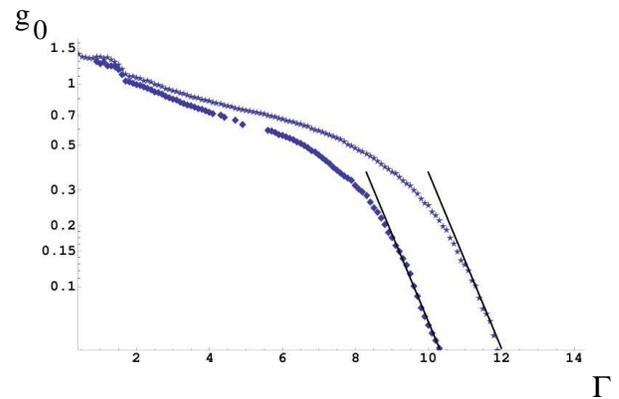}
\caption{A semi-log plot of $g_0$ vs. $\G$ in the Bose glass
  phase. The two samples are the same as the star and diamond curves
  in Fig. \ref{figrmu1} with:
$\delta=-0.02,\,\Delta\on=0.08$ (stars), and
  $\delta=0,\,\Delta\on=0.12$ (diamonds). The dark lines are guide to
  the eye, and have a slope of $-1$ in the plot. As can be seen, the
  late stage of the flow of $g_0$ fits $g_0\sim e^{-\Gamma}$ very
  well.
  \label{gcompfig}}
\end{figure}

We now turn to the analysis of the insulating phase at generic
disorder, i.e. of the Bose glass. Even though the simple exponential
form of $f(\zeta,\on)$ and $g(\beta)$ does not give the exact
solution to the flow equations, as we deduce from numerical
analysis, it gives a very good approximation to the true
distributions. For large values of $f_0$ the equation (\ref{ans2})
simplifies to $f(\zeta,\on)\approx f_0^2 \exp(-f_0 \zeta)$. It is
straightforward to check that under these conditions we have $
f_1\approx 1$ and $f_s\approx 4f_0\exp(-f_0/2)$. Then the flow
e1quation for $g_0$ becomes very simple:
\be
{dg_0\over d\Gamma}\approx -g_0.
\ee
Such flow indicates that the Bose glass phase is indeed intermediate
between the Mott Glass where $g_0^\prime\approx -f_0 g_0$ and the
random singlet insulator with $g_0^\prime\approx -g_0^2$. Physically
the parameter $g_0$ characterizes the strength of the hoppings
remaining in the system. As we argued before slow $1/\Gamma$ decay
of $g_0$ in the random singlet phase resulted in the divergent
density of states at zero energy and as a result in a divergent
compressibility. On the other hand in the Mott glass $g_0$ was
vanishing very rapidly $g_0\sim \exp-(\exp(\Gamma))$ and thus the
corresponding Mott glass had a vanishing density of states at zero
energy and vanishing compressibility. In the Bose glass phase we
have exactly intermediate behavior: $g_0\sim\exp(-\Gamma)$. As we
will see shortly this scaling implies finite density of states at
low energies and thus a finite compressibility. In
Fig.~\ref{gcompfig} we plot the flows of the parameter $g_0$ as a
function of $\G$ obtained from numerical solution of the RG
equations for two different insulating samples with generic disorder. As can be seen from
the latter figure, the flow of $g_0$ vs. $\G$ is consistent with
$g_0\sim \exp(-\G)$.

Another interesting conclusion is coming from the fact that
$f_1\approx 1$  is independent on $f_0$ (and hence
$dg_0/d\Gamma\approx -g_0$). We remind that $f_1 d\Gamma$ gives the
probability of decimating a site when we change the cutoff scale
from $\Omega$ to $\Omega(1-d\Gamma)$. It is remarkable that in the
Bose glass phase this probability is independent of $f_0$. On the
contrary, the probability of eliminating the link is proportional to
$g_0$ and thus vanishes at long $\Gamma$. Thus we come to the
conclusion that the number of the sites remaining in the system $N$
scales exactly as the cutoff energy scale:
\be
N\approx \kappa \Omega,
\label{rhoohm}
\ee
which indicates uniform density of localized states in the insulating state. As we will see shortly the parameter $\kappa$ plays the role of the
compressibility. It is interesting to note that $\kappa$
discontinuously changes across the phase transition. Indeed very
close to the transition the number of sites remaining in the system
is given by
\be
N(\Gamma)\sim N_0\exp(-g_0\Gamma)\sim \exp(-\Gamma).
\ee
This behavior is correct for length scales shorter than the
correlation length $\xi\sim\exp(1/\sqrt{A})$, where $A$ is the tuning
parameter appearing in Eq.~(\ref{crit_flow}), $A\to 0$ corresponding
to the critical point. After that we should use the flow equations
valid in the insulating regime where $N\sim
\exp(-f_1\Gamma)\sim\exp(-\Gamma)$. So we see that the scaling
$N\sim N_0\exp(-\Gamma)$ works very well. We thus conclude that the
ratio $N(\Gamma)/\Omega(\Gamma)$ goes to a constant, which is
independent of $A$.

\subsubsection{Compressibility and SF susceptibility}

The easiest way to see that $\kappa$ is indeed the compressibility
is to map the renormalized array of clusters into a spin-1/2 chain.
Since deep in the insulating phase the displacement $|\on|$ is close
to $1/2$, the local interaction strengths, $U_i$, is mostly quite
large, and obeys: $U_i>\Omega$. This implies that we could retain
only the two lowest charing states, which we the spin up and spin
down states of an effective spin $1/2$ degree of freedom:
\be
n_i=\frac{1}{2}+\hat{s}^z.
\ee
In this picture the gap $\Delta_i=U(1-2|\on|)$ plays the role of the
external magnetic field along the $z$ axis $h^z_i$. The hopping
$J_i$ is in turn maps to the $xy$ coupling between the neighboring
spins.

Let us first determine the distribution function of $h^z$: $H(h^z)$
assuming that $f(\zeta,\on)$ is given by Eq.~(\ref{ans2}) with
$f_0\gg 1$.
\beq
&& H(h^z)\approx f_0^2 \intt_{-1/2}^{1/2}\!\!\!
d\on\!\!\!\intt_{1-2|\on|}^{\infty}\!\!\!d\zeta\, \mathrm
e^{-f_0\zeta}
\delta\l(|h^z|-\frac{\Omega}{\zeta}(1-2|\on|)\r)\nonumber\\
&&~~~~~~~~\approx {1\over 2\Omega}\,\chi(\Omega-|h^z|),
\eeq
which is just a uniform distribution.

The compressibility of the insulating phase is given by the z-field
susceptibility of the spin chain. The latter is easily shown to be
twice the probability density of $H(h^z=0)$. Thus the
compressibility is:
\be
\kappa=\der{n}{\mu}=N\der{s^z}{h^z_{\rm
ext}}=2N{H(h^z=0)}=\frac{N}{\Omega},
\ee
where $h^z_{\rm ext}$ is the infinitesimal external magnetic field
along the $z$ axis. Indeed, this is the result advertised in Eq.
(\ref{rhoohm}).

The superfluid susceptibilty $\chi_s$ is obtained as the response of
the spin chain on a small magnetic field in $x$ direction. Note that
in the Bose glass phase the coupling between different sites is
vanishingly small, and thus $\chi_s$ can be derived by considering an
isolated site, which is described by a spin $1/2$ Hamiltonian:
\be
\H_i=h^z\hat{s}^z_i+h^x_{\rm ext}\hat\acute{}{s}^x.
\ee
A straightforward calculation yields that the average magnetization
along the $x$ axis is:
\be
\overline{\langle
 \hat{s}^x\rangle}={1\over 2}\overline{\frac{h^x_{\rm ext}}{\sqrt{(h^x_{\rm
 ext})^2+(h^z)^2}}}.
\ee
Thus the susceptibility is:
\be
\chi_{s}=N \intt_{|h^x_{\rm
ext}|}^{\Omega}\frac{dh^z}{2\Omega}\frac{1}{|h^z|}=\frac{N}{2\Omega}\log
\frac{\Omega}{|h^x_{\rm
ext}|}=\frac{\kappa}{2}\log\frac{\Omega}{|h^x_{\rm ext}|}.
\ee
Obviously $\chi_s$ diverges as $h^x_{\rm ext}\to 0$. Thus we find
that the Bose glass phase is characterized by divergent $\rho_s$ and
finite $\kappa$ in agreement with Ref.~[\onlinecite{FWGF}].

\section{Compressibility, stiffness, and Luttinger parameter in the
  superfluid phase and at criticality \label{SFLutt}}

Let us now focus on the properties of the superfluid
phase we find. The superfluid phase is associated with the formation
of a superfluid cluster that spans the chain. The
cluster consists of all the original bare sites that were
not decimated due to their charging energies. These surviving sites
obey the Hamiltonian:
\be
\H_{eff}^{SF}=\summ_i \l[\frac{1}{2}\tilde{J}_i\l(\f_{i+1}-\f_i\rr)^2+U_i \hat{n}_i^2\rr]
\label{SFeff}
\ee
where $\f_i$ and $\hat{n}_i$ are the phase and number operators of the
surviving cites, and $U_i$ are the charging energies of each bare
site. The $\tilde{J}_i$ are harmonic couplings
between the surviving sites, which are the result of the decimation of
a strong bond (marked with a tilde since they can get renormalized by
intervening charge-blockaded sites) as we now explain. When deriving the RG equations,
we eliminated the strongest bonds iteratively, by setting the sites
they connected into phase-coherent clusters, which implies replacing
the strongest Josephson couplings with a harmonic coupling:
\be
-J_i\cos\l(\f_{i+1}-\f_i\rr)\rightarrow\frac{1}{2}J_i
\l(\f_{i+1}-\f_i\rr)^2.
\label{approJ}
\ee
We then approximated the cluster to be phase-coherent:
\be
\tilde{\f}\approx\f_i\approx\f_2.
\ee
This strong approximation is sufficient for obtaining the flow
equations, but we need to allow intra-cluster fluctuations it in order to discuss the
properties of the superfluid phase.

The stiffness and the compressibility of the superfluid phase are
given in terms of the parameters $\tilde{J}_i$ and $U_i$ in the
effective Hamiltonian (\ref{SFeff}), which describes the proliferating superfluid cluster.
The compressibility is given by
\be
\kappa=\frac{1}{L}\frac{1}{U_{SF-cluster}} =\frac{1}{L}\summ_{i\in SF} \frac{1}{U_i}
\label{eq1}
\ee
where $L$ is the total length of the chain. The inverse superfluid stiffness is similarly obtained as
\be
\frac{1}{\rho_s}=\frac{1}{L}\summ_{i \in SF} \frac{1}{\tilde{J}_i}.
\label{eq2stiffness}
\ee
Note that the simple expression for the stiffness owes to the fact that the fixed point Hamiltonian (\ref{SFeff}) is harmonic. Therefore the stiffness suffers no further renormalization
by quantum fluctuations and it is the same as in the classical model (see [\onlinecite{AKPR}]).

We will now proceed to calculate the average compressibility , stiffness,
and Luttinger parameter of the superfluid $K\equiv \pi\sqrt{\kappa\rho_s}$ .

\subsection{Differential equation for the inverse charging energy}

The compressibility given by Eq. (\ref{eq1}) can be calculated in a rather
straight forward way within the RG scheme outlaid in the previous sections.
The variable $\zeta$ in the RG scheme is specifically
designed to keep track of the cluster compressibilities.
We recall the RG flow equation (\ref{master}) for the distribution function $f(\zeta)$
\be
\ba{c}
\frac{df(\zeta)}{d\G}=(1+\zeta)\der{f(\zeta)}{\zeta}\vspace{2mm}\\
+g_0\int d\zeta_1d\zeta_2 \delta(\zeta-\zeta_1-\zeta_2-1)\vspace{2mm}\\
f(\zeta_1)f(\zeta_2)
+f(\zeta)(f_0+1-g_0).
\ea
\label{diff2}
\ee
The solution to this equation
will allow us to compute the compressibility from the average value of $\zeta$ as
$\kappa={\bar \zeta}/\Omega$.

%This equation was shown to have an attractor in the form of
%Eq. (\ref{scaling}). In principle, this allows us to calculate the
%full distribution of the compressibility, once we know at which stage
%of the RG to stop counting the inverse charging terms.

To obtain a differential equation directly for the
average compressibility (inverse-charging energy) we move to the
Laplace-transformed representation:
$F(\eta)=\int_0^{\infty} e^{-\eta\zeta}f(\zeta)$, which obeys:
\be
\ba{c}
\frac{dF(\eta)}{d\G}=-f_0+F(\eta)(\eta-1)-\eta\der{F(\eta)}{\eta}\\
+g_0 F(\eta)^2e^{-\eta}+F(\eta)(f_0+1-g_0).
\ea
\label{diff3}
\ee
We now use that:
\be
\overline{\zeta}=-\l.\der{F(\eta)}{\eta}\r|_{\eta\rightarrow 0}
\ee
to obtain:
\be
\frac{d\overline{\zeta}}{d\G}=\overline{\zeta}(f_0+g_0-1)-1+g_0
\label{Udiff1}
\ee

The inverse charging energy is given by Eq. (\ref{eq1}), in which an
extra factor of $\O$ appears. Adding it on we obtain:
\be
\frac{d\gamma}{d\G}=\gamma(f_0+g_0)-(1-g_0)/\O
\label{Udiff}
\ee
where $\gamma=\overline{\zeta}/\O$.

\subsection{Flow equation for the stiffness}

Calculation of the stiffness requires a slight extension of the RG scheme.
The method described thus far did not include a cluster variable
which stores the internal stiffness. In other words the RG scheme does not keep track
of the internal sum over $1/J_i$ (\ref{eq2stiffness}) within the proliferating clusters.

Fortunately, such a variable can easily be easily included by extending the cluster distribution
function $f(\zeta)$ to a joint distribution $f(\zeta,\chi)$, where
\be
\chi=\summ_{i \in cluster} \frac{1}{\tilde{J}_i}
\ee
is a variable
designed to keep track of the superfluid stiffness of the clusters. Each time two
clusters are joined in the RG flow by a large bond $J=\Omega$,
the variable $\chi$ of the joined
cluster is given by:
\be
\chi_{12}=\chi_1+\chi_2+\frac{1}{\O}=\chi_1+\chi_2+\frac{1}{\O_0}e^{\G}.
\ee

The flow equation for $f(\zeta,\chi)$ is a straight forward extension of Eq. (\ref{diff2}):
\begin{widetext}
\be
\ba{c}
\frac{df(\zeta,\chi)}{d\G}=(1+\zeta)\der{f(\zeta,\chi)}{\zeta}\vspace{2mm}\\
+g_0\int d\zeta_1d\zeta_2 \delta(\zeta-\zeta_1-\zeta_2-1)\int
d\chi_1d\chi_2 \delta(\chi-\chi_1-\chi_2-\O_0^{-1} e^{\G})
f(\zeta_1,\chi_1)f(\zeta_2,\chi_2)\vspace{2mm}\\
+f(\zeta,\chi)(f_0+1-g_0).
\ea
\label{diff1}
\ee
\end{widetext}
This is a rather complicated equation for the joint distribution of
cluster stiffness and charging energy. However it can be
greatly simplified if we are interested only in the average of the stiffness.
The latter can be calculated by integrating Eq. (\ref{diff1}) with respect to $\zeta$, and taking its Laplace transform with respect to $\chi$:
$S(\lambda)=\intt_0^{\infty}d\zeta \intt_0^{\infty}d\chi e^{-\lambda\chi}f(\zeta,\chi)$. This yields:
\be
\ba{c}
\frac{dS(\lambda)}{d\G}=-\tilde{f}(0,\lambda)-S(\lambda)\vspace{2mm}\\
+g_0S(\lambda)^2e^{-\lambda\cdot\frac{1}{\O_0}
  \exp \G}+S(\lambda)(1+f_0-g_0)
\ea
\label{Jdiff1}
\ee
Where:
\[
\ba{cc}
\tilde{f}(0,\lambda)=\int d\chi e^{-\lambda\chi}f(\zeta=0,\chi)
\ea
\]
Again using the fact that:
\be
\overline{\chi}=-\l.\der{S(\lambda)}{\lambda}\r|_{\lambda\rightarrow 0}
\ee
we obtain:
\be
\frac{d\overline{\chi}}{d\G}=\overline{\chi}(f_0+g_0)+g_0/\O+\l.\der{\tilde{f}(0,\lambda)}{\lambda}\r|_{\lambda\rightarrow 0}
\label{Jdiff2}
\ee
where we used $\O=\O_0\exp(-\G)$
We note that the only difference between Eq. (\ref{Udiff}) and
Eq. (\ref{Jdiff2}) is in the subleading term, $g_0/\O$ above, and $(g_0-1)/\O$ in
Eq. (\ref{Udiff}). There is also the last term in Eq. (\ref{Jdiff2}),
which should be negligible and negative.

\subsection{Differential equation for the length of a superfluid
  cluster}

The differential equation for the typical length of the clusters\cite{AKPR}
is given by
\be
\frac{d\ell}{d\G}=\ell(f_0+g_0)
\label{lengtheq}
\ee
It is interesting to note that this equation is the same, at the leading order,
as the equations, derived above, for the sums of inverse charging energies $\sim \bar\zeta$
and the sum of inverse Josephson couplings $\bar\x$ within a cluster.

When calculating the stiffness and compressibility using the flow
equations for a particular cluster, as illustrated above, we need to
renormalize until the size of a SF cluster is that of the entire
chain:
\be
\ell_{\G}=L.
\ee
Therefore the compressibility:
\be
\kappa=\frac{1}{L}\frac{1}{U_{SF-cluster}}
=\frac{\overline{\zeta}/\O}{\ell_{\G}}
\ee
and the inverse stiffness:
\be
\frac{1}{\rho_s}=\frac{1}{L}\summ_{i \in SF}
\frac{1}{J_i}=\frac{\overline{\chi}}{\ell_{\G}}
\ee
always tend to a number as $\G\rightarrow \infty$.

\subsection{Compressibility and stiffness at the critical point}

Let us assume that we start sufficiently close to the critical point, so that the distributions $f(\zeta,\G)$ and $g(\beta,\G)$ already converged to the universal forms  characterized by $f_0(\G)$ and $g_0(\G)$ (see Eqs.~(\ref{ans1}) and (\ref{ans2})).
In Sec. (\ref{FPscale}) we found the explicit flow of these functions at criticality
$g_0-1\approx 2/\G+2/\G^2$ and $f_0\approx 4/\G^2$ (that is, the flow on the separatrix).
These flows start at some initial value $\G_0$ which characterizes the bare disorder distributions of the microscopic system. The larger is $\G_0$ the wider is the disorder distribution in $J_i$.

Combining this with flow equation for $\zeta$ (\ref{Udiff1}) we find
\[
\frac{d\overline{\zeta}}{d\G}=\overline{\zeta}\left(\frac{2}{\G}+\frac{6}{\G^2}\right)+\frac{2}{\G}.
\]
This has the solution for large $\G$:
\be
\overline{\zeta}\approx C_{\zeta}\cdot \G^2\exp[-6/\G]
\ee
with $C_{\zeta}$ being a constant, which we obtain from initial conditions. We know that for sufficiently large $\G_0$, $\overline{\zeta}_0=\frac{1}{f_0^{(0)}}=\frac{1}{4}\G_0^2$. This
implies:
\be
\overline{\zeta}=\frac{1}{4}\G^2\exp[-6/\G]
\ee
and $C_{\zeta}=1/4$.

Similarly, for $\overline{\chi}$ we get:
\be
\overline{\chi}\approx \frac{1}{\O} {1\over 6}\G^2\exp[-6/\G]\l(C_{\chi}-\exp[6/\G]\r),
\ee
where the second term in the brackets comes from the $g_0$ term in the equation for $\chi$. The constant $C_{\chi}$ can also be obtained from boundary conditions. At the onset $\overline{\chi}_0=0$ since we start with bare sites, and only after some RG we get the sum of $1/J$ to grow. This implies:
\be
C_{\chi}=\exp[6/\G_0].
\ee

By the same token, the solution of Eq. (\ref{lengtheq}) at criticality
is:
\be
\ell_{\G}=\ell_0 e^{\G}\frac{\G^2}{\G_0^2}
\ee
where $\ell_0$ is of order 1.

Putting our results in the definitions of $\rho_s$ and $\kappa$, we
obtain the compressibility of the critical system sending $\G\to\infty$:
\be
\kappa=\frac{\overline{\zeta}/\O}{\ell_{\G}}
=\frac{\G_0^2}{4\O_0\ell_0}
\label{compres}
\ee
and the inverse stiffness:
\be
\frac{1}{\rho_s}=\frac{\overline{\chi}}{\ell_{\G}}
=\frac{\G_0^2}{6\O_0\ell_0}(\exp[6/\G_0]-1).
\label{stiffres}
\ee
The energy scale for both the stiffness and the compressibility is
given by $\O_0$, the initial energy scale of the problem. Both also
tend to constants along the critical flow line.

\subsection{Luttinger parameter at criticality }

By multiplying  Eqs. (\ref{compres}) and (\ref{stiffres}) we obtain
the luttinger parameter of the SF cluster:
\be
K^2=\pi^2 \kappa\rho_s={3\pi^2\over 2}{1\over \exp[6/\G_0]-1}.
\label{Luttres}
\ee
Indeed we find that it is a constant along flows on the critical
manifold, which is independent of the initial energy scale $\O_0$.
On the other hand this result is clearly not universal, since it depends on
$\G_0$.

As mentioned above, $\G_0$ parameterizes the strength of the bare bond disorder distribution. For a given system on the critical manifold, the larger is $\G_0$ the broader is the
system's initial distribution of both $J$ and $1/U$. We can therefore interpret
Eq. (\ref{Luttres}) as stating that at strong disorder, the Luttinger parameter required
to stabilize a superfluid phase depends on the disorder strength. A larger Luttinger parameter
is needed the more disordered is the system.
This statement is clearly different from the situation at weak disorder, for which
Giamarchi and schulz had predicted a transition at a universal value of the Luttinger parameter\cite{Giamarchi-Schulz1987,Giamarchi-Schulz1988}. We shall comment on the relation between these two limits in the discussion below.

\section{Discussion \label{discussion}}

Using the real-space RG approach, we obtain a consistent picture both
of the possbile insulating phases of the random-Bose-Hubbard model,
but also of the transition from the superfluid to them. The seminal
work of Giamarchi and Schulz (GS)\cite{Giamarchi-Schulz1987,Giamarchi-Schulz1988} obtained a description of what seems to be the same
transition in terms of a perturbative RG in weak-randomness - the
opposite limit to our starting point. We now ask: how do these two
scenarios, or descriptions, correspond to each other?
Now that we obtained our result for the Luttinger parameter at criticality,
Eq. (\ref{Luttres}), we can address this question.

One of the central results of Ref.~\cite{Giamarchi-Schulz1988} is the
universality of the Luttinger parameter at the transition:
\be
K^{(GS)}_c=\frac{3}{2}.
\ee
Since GS considered the anomalous dimension of what is essentially a
phase-slip operator, the universality of $K$ at the transition was
deduced from the fact that when $K>3/2$, phase slips are
irrelevant. Since in weak randomness phase slips are clearly the most
relevant operators, the vanishing of their scaling dimension implies
criticality. Also, the generality of the GS approach, and the
self-averaging of the SF phase~\cite{Prokofiev-Private} implies that phase slips turn relevant when $K=3/2$ even for strong disorder.

At strong randomness, however, we find that a different type of
disturbance of the superfluid phase can disorder it. In the real-space RG analysis grains with large charging energies are decimated, implying that a whole grain becomes isolated from the rest of the chain. This
process is equivalent to a phase-slip dipole happening around the
grain. Phase-slip dipoles consist of a phase slip and an anti-phase
slip happening simultaneously at neighboring positions in the
chain. In the week coupling limit, these dipoles are not enough to
degrade the superfluidity, since they do not produce a voltage
drop. But when the disorder is strong, the dipoles, or equivalently,
the blockaded insulating sites, suppress tunneling across
the lattice, as we find from our analysis.

For sufficiently strong disorder, the Luttinger parameter at which
blockaded sites destroy superfluidity, i.e., the critical Luttinger parameter, is given by
Eq. (\ref{Luttres}):
\be
K=\pi\sqrt{{3\over 2}\frac{1}{\exp[6/\G_0]-1}}.
\ee
For the p-h symmetric case considered in Sec. \ref{phs}, and
$K=\pi\sqrt{2/[\exp(4/\G_0)-1]}$ for the commensurate case, with
$\overline{n}_j=0$. As explained below Eq. (\ref{Luttres}), $\G_0$ is a measure of the initial disorder of the system. Thus, $K$ grows monotonically with the disorder, and exceeds the universal GS value of $K=3/2$ at intermediate values of $\G_0\sim 3$. This implies that the transition we find takes over the universal GS transition at a finite disorder: since we find that the breakdown of superfluidity occurs at $K>3/2$,the transition happens well into the region where single phase-slips are irrelevant, and thus they do not modify the critical properties of the model, and can be safely ignored. This also justifies our procedure of SF cluster formation as outlined in Eq. (\ref{approJ}). It is interesting to note that $\G_0\sim 3$ corresponds to a charging distribution which is peaked at about $4/\G_0^2 \Omega_0\sim 1/2\Omega_0$, as obtained by plugging $\G_0$ into Eqs. (\ref{scaling}).

\begin{center}
\begin{figure}
\includegraphics[width=8cm]{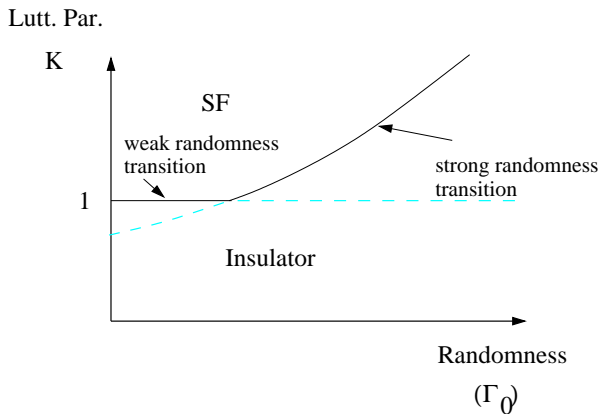}
\caption{From all the analyses we carried out it seems that our
  transition does not happen at a universal value of the luttinger
  paerameter, but rather, at a universal value of the power-law of the
  J distribution ($g_0=1$). From the discussion, it seems that there
  are two scenarios for the breakdown of the SF. At weak randomness it
  is the G-S single-vortex proliferation mechanism that first
  destabilizes the SF. In this range, the scenario we present would destabilize
  the SF at lower $K$ than single vortex proliferation (dashed grey), and is
  therefore not a true boundary.  At larger randomness, our scenario
  is the first to stabilize the SF, as it occurs at larger K's than
  the universal G-S value. A concequence is that the universality of
  the Luttinger parameter at criticality is lost. \label{prPD}
}
\end{figure}
\end{center}

Our conclusion is that at finite randomness the critical fixed
point of the RSRG takes over (Fig. \ref{prPD}). When this happens,  universality of the Luttinger parameter at the transition is lost. Since the transition we are describing is still a Kosterlitz-Thouless type transition, many properties of the weak-randomness transition, and strong randomness
transition are shared. One can argue that the Luttinger-parameter
universality lost at strong disorder morphs into a different
universlity --- that of the exponent with which the distribution of
$J$ vanishes at small energies, which is $g_0-1\rightarrow 0^+$ at
criticality.

An outstanding question is how the weak-randomness phase-slip
driven transition changes into the transition we find at
strong disorder. One possiblity is that the two scenario continuously
morph into each other. Yet another more exciting possiblity is that
our analysis is equivalent to the calculation of the scaling-dimension of an operator different from single phase-slips, and that such an operator becomes relevant at sufficiently strong disorder at Luttinger parameters $K>3/2$.
Therefore it causes a break down of superfluidity before phase-slips become
relevant.

Another important difference between the perturbative approach of GS and our results is that GS assume that the diagonal disorder is gaussian and fully characterized by its variance while the off-diagonal disorder is weak and irrelevant. On the contrary, in the strong randomness approach, we see that transition corresponds to a wide power-law distribution of tunneling amplitudes. Standard replica methods are not applicable to this type of disorder distribution and thus it is not surprising that the jump we find in $K$ is different. The appearance of broad power law distribution of links in 1D is not surprising. There is always a finite chance of encountering a large insulating cluster separating two superfluid regions, which effectively blocks the tunneling between superfluids. This is a special property of 1D systems. In Ref.~\cite{AKPR} we demonstrated that this is indeed the case for a simple toy model. The real space RG just reflects this property of 1D systems. Thus if disorder is not very small so that weak links necessarily occur with finite probability, we believe that our scenario of the SF-IN transition to be more plausible than GS scenario of weak disorder. However, the final resolution of this question is currently beyond reach of both the RSRG and the GS analysis, since it is concerned with the intermediate randomness regime. Probably this question can be addressed numerically.

\section{Experimental concequences}

\subsection{Critical current of a finite superfluid chain}

At low energy scales, the real-space renormalization group allows
detailed knowledge of the superfluid phase. Most importantly, the
effective low-energy Josephson junction coupling distribution is:
\be
\rho(J)=g_0 \frac{1}{J^{1-g_0}}.
\label{plj}
\ee
The knowledge of the Josephson distribution function allows us to make
a connection with a rather simply measurable experimental property:
The critical current of a chain.

Unlike the Luttinger parameter, the critical current of a bosonic
chain in the absence of phase fluctuations is controlled by its
weakest hopping link. Given a strongly disordered bosonic chain in its
superfluid phase, we can apply the real-space RG until the effective
coupling distirbutions approach their universal behavior, and in
parituclar the distribution (\ref{plj}) for the Jospehson energies of
each bond, and with negligible charging effects.

Let us now calculate the scaling of the critical current on the bare
length of the system.  Given a particular disorder distribution, the universal
distributions are obtained once the UV cutoff is $\Omega_0$, and only a fraction  $1/\ell$ of the
chain is still active, and the chain is of length $L/\ell$. The
scaling behavior of the weakest Josephson energy expectation value,
$J_{min}$,  is obtained by requiring that the probability of having at
least one
bond with an energy $J<J_{min}$ is of order 1, which translates to the condition:
\be
\frac{L}{\ell} \intt_{0}^{J_{min}}\frac{dJ}{\Omega_0}\frac{g_0}{(J/\Omega_0)^{1-g_0}} =1
\ee
Carrying out the integral we obtain:
\be
J_{min}\sim \l(\frac{\ell}{L}\rr)^{1/g_0}.
\label{Jscale}
\ee
In the weak disorder regime, where $g_0\gg 1$ we see that the critical current is almost size independent. While at strong disorder near the transition $g_0\to 1+$ the critical current scales as the inverse system size. This prediction can be directly tested in experiments. Using extreme value statistics one can even find the whole Gumbel distribution of the critical current in the SF regime:
\be
P(J_{min})\sim {g_0\over J_{min}^{1-g_0}}\exp\left[-{L\over \ell} (J_{min})^{g_0}\right].
\ee

\subsection{Resistance at finite temperatures}

By a similar argument, we can guess the finite temperature behavior of
a disordered superfluid chain. First, we make the following
simplifying assumptions: if a bond strength is $J>T$, we can neglect its
finite temperature resistance, but if $J<T$, a bond will give a finite
resistance $r$, which is $T$ independent. Furthermore, we ignore, for
the sake of this discussion, the dependence of $r$ on $J$.

Under these simple assumptions, the resistivity $\rho$ at temperature $T$ is
given by the density of bonds of strength $J<T$. Therefore:
\be
\rho\sim
r\int_0^{T}\frac{dJ}{\Omega_0}\frac{g_0}{(J/\Omega_0)^{1-g_0}}=r
(T/\Omega_0)^{g_0}
\label{RT}
\ee
where, as defined above, $\Omega_0$ is the rough energy scale at which
the chain is exhibiting the universal low energy behavior.

In finite chains, we expect that Eq. (\ref{RT}) would only be valid
when $T\gg J_{min}$. Very crudely, by replacing the lower limit of the
integral in Eq. (\ref{RT}) by $J_{min}$ as given by Eq. (\ref{Jscale}), we obtain for $T>J_{min}$:
\be
\rho\sim r \l|\l(\frac{T}{\Omega_0}\r)^{g_0}-\frac{\ell}{L}\r|.
\ee

\section{Conclusions}

In this paper we extend the real-space RG analysis of
Ref. [\onlinecite{AKPR}] to the case of non-commensurate chemical
potential. We find that remarkably, the symmetry and details of the
diagonal disorder are irrelevant for the SF-INS transition in a system
with only onsite interactions. Nevertheless, the symmetry of the
disorder completely determines the type of insulator that the system
obtains. The superfluid phase will break down at a Kosterlitz-Thouless
critical point, and will become: (i) a gapless, incompressible,
Mott-glass if the chemical potential is commensurate
($\overline{n}_j=0$), (ii) a gapless, compressible Bose-glass with
diverging superfluid susceptibility if $1/2<\overline{n}_j\le 1/2$ is
unrestricted, (iii) a gapless random-singlet glass with a diverging
compressiblity and superfluid susceptibilty in the case of p-h
symmetric chemical potential ($\overline{n}_j=0,\,1/2$).

An important question about our approach is its connection with the seminal work of Giamarchi and Schulz~\cite{Giamarchi-Schulz1987}, we
calculated the properties of the superfluid phase using the real-space
RG analysis. By considering the Luttinger-parameter $K$, we showed that at
strong disordered the SF-INS transition occurs at a finite value of
$K$, larger than the universal GS value, and that the universality of the Luttinger parameter is replaced with a universality of the power-law
distribution of effective hopping at low energies. The real-space RG
approach is thus not complementary to the GS approach, but provides a
description of the SF-INS transition at strong disorder, and allows
direct access to the insulating phases, where the GS approach fails.

An interesting direction to pursue in the future is the utilization
of the RSRG approach for calculation of transport-propoerties, and
finite-temperature properties of the random 1-d Bose-Hubbard
chain. Such calculations could probably be done by combining our
approach with that of Motrunich, Damle, and Huse \cite{MotrunichDamleHuse}. The presence of very large disorder in the insulating phases should make
such calculations accessible. On the other hand, they may prove
difficult near the transition due to the finite randomness there.

Another outlying question is that of the correlations in the SF
phase. Self averaging indicates that the Luttinger parameter we find
in Sec.~\ref{SFLutt} also dictates the decay of correlations in the
strongly-disordered superfluid phase. This, however, remains to be
confirmed in direct numerical investigation of a strongly random
harmonic chain. We should emphasize that due to our method for finding the
Luttinger parameter, it should be consistent with the anomalous
dimension of the phase-slip operator in the GS theory.

{\em Acknowledgments.} We are most grateful to  S.~Girvin for the useful
suggestion to look into the half integer case first, and to
D.S. Fisher, M.P.A. Fisher, T. Giamarchi, V. Gurarie, P. Le-Doussal, O. Motrunich, N. Prokofe'v, B. Svistunov
for numerous discussions. A.P. acknowledges support from AFOSR YIP and
Sloan foundation, G.R. acknowledges support of the Packard foundation,
Sloan foundation and the Research Corporation, as well as the
hospitality of the BU visitor program.

\bibliography{random}
\end{document}